\documentclass{ws-mpla}
\usepackage{amssymb}

\renewcommand{\bar}[1]{\overline{#1}}

\providecommand{\Journal}[4] {#1 {\bf #2} (#4) #3}
\providecommand{\EPJA}{Eur. Phys. J. A } %
\providecommand{\MPLA}{Mod. Phys. Lett. A} %
\providecommand{\NPA}{Nucl. Phys. A } %
\providecommand{\NPB}{Nucl. Phys. B } %
\providecommand{\PL}{Phys. Lett. } %
\providecommand{\PLB}{Phys. Lett. B } %
\providecommand{\PRL}{Phys. Rev. Lett. } %
\providecommand{\PRC}{Phys. Rev. C } %
\providecommand{\PRD}{Phys. Rev. D } %
\providecommand{\PRSA}{Proc. Roy. Soc. A } %
\providecommand{\ZPA}{Z. Phys. A } %

\begin{document}

%

\def\nocropmarks{\vskip5pt\phantom{cropmarks}}

\let\trimmarks\nocropmarks      

%

\markboth{Xun Chen, Yajun Mao, Bo-Qiang Ma}
{Decay Probability Ratio of Pentaquark $\Theta^+$ State}

%
\catchline{}{}{}{}{}
%

\title{DECAY PROBABILITY RATIO OF PENTAQUARK $\Theta^+$ STATE  }

\author{\footnotesize XUN CHEN}

\address{School of Physics, Peking University, Beijing 100871, China}

\author{YAJUN MAO}

\address{School of Physics, Peking University, Beijing 100871, China}

\author{BO-QIANG MA\footnote{
Corresponding author. Email address: mabq@phy.pku.edu.cn}}

\address{
School of Physics, Peking University, Beijing 100871, China
}

\maketitle

\pub{Receivd (Day Month Year)}{Revised (Day Month Year)}

\begin{abstract}
The pentaquark state of $\Theta^{+}(uudd\bar{s})$ has been
observed to decay with two decay modes: $\Theta^+\to n K^+$ and
$\Theta^+ \to p K^0$. The decay probability ratio of the two decay
modes is studied with general symmetry consideration of isospin,
spin, and parity. We arrive at a result of the ratio
$\frac{\Gamma(\Theta^+\to nK^+)}{\Gamma(\Theta^+\to pK^0)}
=\frac{(\alpha-\beta)^2}{(\alpha+\beta)^2}(\frac{k_1}{k_2})^{2L+1}$,
which is valid for the $\Theta^+$ state to be a pure isoscalar or
isovector state, or an isoscalar and isovector mixing state, or an
isotensor state with mixture of isoscalar and isovector components
with coefficients $\alpha$ and $\beta$. The dependence on spin and
parity of the pentaquark $\Theta^+$ state is found to be small due
to small difference between the center of mass decay momenta $k_1$
and $k_2$ of the two decay modes. We also provide an analysis on
the constraint of the isospin of $\Theta^+$ from the absence of a
peak in the $pK^+$ invariant mass distribution in the $\gamma p
\to pK^+K^-$ process. Future experimental results about the decay
probability ratio may  provide information about the properties of
the pentaquark $\Theta^+$ state.

\end{abstract}

\ccode{PACS Nos.: 12.39.Mk; 11.30.-j; 11.30.Er; 13.30.Eg}

The standard quark model\cite{QM} is remarkably successful in
classifying hadrons as composite systems of quark-antiquark
($q\bar{q}$) states for mesons and three quark $(qqq)$ states for
baryons. The possible existence of multiquark states with four
quarks and an antiquark, i.e., the pentaquark states with
configuration $qqqq\bar{q}$,\cite{qqqqqbar,pentaC,GM99} has been
discussed for more than two decades. QCD based models permit the
existence of pentaquark states.\cite{qqqqqbar}
The pentaquark states with the antiquark $\bar{q}$ being a heavy
quark are suggested to exist. \cite{pentaC} Earlier experimental
attempts were focused on searching for such pentaquark states
containing at least one heavy flavor quark (antiquark), but
without convincing evidence.\cite{Lip97} It is noticed that
$qqqq\bar{q}$ is unambiguously a minimal pentaquark state if the
flavor of $\bar{q}$ is different from any of the other four
quarks.\cite{GM99}
 The possible existence of the minimal pentaquark
states with the antiquark being an anti-strange quark $\bar{s}$,
such as $\Theta^{+}(uudd\bar{s})$ and $\Theta^{++}(uuud\bar{s})$,
are discussed in Ref.~\refcite{GM99}, with suggestion for
experimental searches also provided.

 From another point of view, baryons can be viewed as solitons in
an effective meson theory\cite{Skyrme} as well as in the large
$N_c$-limit of QCD.\cite{Witten} Thus baryons can be also
classified by soliton configurations in chiral field theories, as
an alternative scheme compared to the standard quark model
classification. In the SU(3) version of these models, the
anti-decuplet $\{\bar{10}\}$ baryon multiplet is the lowest
multiplet above the minimal octet $\{8\}$ and decuplet $\{10\}$
baryons.\cite{Mano-Chemtob}  It is
found\cite{Pra,penta1,penta2,penta3} that there exists a baryon
state with strangeness $S=+1$, which can be identified with the
pentaquark $\Theta^+(uudd\bar{s})$ configure, in this
$\{\bar{10}\}$ multiplet. By treating the experimentally observed
N(1710) resonance as a member of this multiplet, Diakonov, Petrov,
and Polyakov predicted\cite{penta2} the $\Theta^+$ baryon state to
have a mass of about $1530~\mbox{MeV}$ and a width of about
$15~\mbox{MeV}$. This baryon state is nothing exotic compared to
other baryons in the soliton classification scheme, except that it
has a strangeness number $S=+1$ and has also a surprising narrow
width.

The recent observations of an exotic narrow baryon state with
$S=+1$ by LEPS,\cite{SPring} DIANA,\cite{DIANA}  CLAS, \cite{CLAS}
SAPHIR,\cite{SAPHIR} and HERMES\cite{HERMES} Collaborations,
provide evidence for the existence of the pentaquark state
$\Theta^+(uudd\bar{s})$. The detection of the $\Theta^+$ state in
LEPS and CLAS experiments is through the subprocess $\gamma n \to
K^- (K^+ n)$, where $K^-$ and $K^+$ are detected with the
$\Theta^+$ state constructed from the missing mass of the $K^+n$
system,\cite{footnote} in a same principle to the $\gamma^* n \to
K^- \Theta^+$ process suggested in Ref.~\refcite{GM99}. The SAPHIR
experiment detects the $\Theta^+(uudd\bar{s})$ state through the
process $\gamma p\to n K_s^0 K^+$.  In DIANA and HERMES
experiments, the $\Theta^+$ state is detected through the decay
mode $\Theta^+ \to K^0 p$, where the $K^0$ meson is constructed by
the decay $K^0_s \to \pi^+ \pi^-$. All of the experiments have
observed a sharp peak in the $K^+ n$ missing mass spectrum or in
the $K_s^0 p$ invariant mass spectrum, with a mass around
$1540~\mbox{MeV}$ and a narrow width around $9\to 25~\mbox{MeV}$,
in agreement with the $\Theta^+$ state predicted by the chiral
quark soliton model.\cite{penta2} Therefore two decay modes:
$\Theta^+ \to  nK^+$ and $\Theta^+ \to p K^0$, have been observed
for the $\Theta^+$ state.

The properties of the observed $\Theta^+$ state, such as its spin,
isospin, and parity, are still not established yet. The chiral
quark soliton model predicts this $\Theta^+$ state of having spin
1/2, isospin 0, and parity + of the anti-decuplet.\cite{penta2}
However, with a mass of $100~\mbox{MeV}$ above threshold, this
pentaquark state should be expected to have a decay width of the
order of $500~\mbox{MeV}$ unless its decays are suppressed by
phase space, symmetry, or special dynamics.\cite{isotensor} It is
suggested by Capstick, Page, and Roberts,\cite{isotensor} that
the narrow width of the observed $\Theta^+$ state can be
understood if the isospin of this state is hypothesized to be 2,
i.e., an isotensor resonance. For this isospin, the decays of
$\Theta^+ \to  n K^+$ and $\Theta^+ \to  p K^0$ via isospin
violating hadronic processes are suppressed by symmetry. Jaffe and
Wilczek tried to explain the narrow width by proposing the
$\Theta^+$ state, which is an isospin singlet, as being composed
of two highly correlated $ud$-pairs and an $\bar{s}$ quark.\cite{JW03}
This work intends to study the probability ratio of
the two decay modes, and to show that this ratio can provide
useful information on the properties of the observed $\Theta^+$
state.

Our analysis is based on general symmetry consideration of isospin
($I,I_z$), spin $J$, and parity $P$. The corresponding quantum
numbers of quarks are
\begin{equation}
(I,I_z)\left(J^{p}\right) = \left\{
\begin{array}{lll}
\left(\frac{1}{2},+\frac{1}{2}\right)\left(\frac{1}{2}^{+}\right),
& \mbox{for the $u$ quark}; \\
\left(\frac{1}{2},-\frac{1}{2}\right)\left(\frac{1}{2}^{+}\right),
& \mbox{for the $d$ quark}; \\
\left(0,0\right)\left(\frac{1}{2}^{+}\right), & \mbox{for the $s$
quark}.
\end{array}
\right.
\end{equation}
The pentaquark $\Theta^+(uudd\bar{s})$ state should have isospin
$I=0$, 1 or 2 with $I_z=0$, therefore we can obtain the general
isospin configuration of this state
\begin{equation}
|\Theta^+ \rangle = a|I=2,I_z=0\rangle +
\alpha|1,0\rangle+\beta|0,0\rangle, \label{sstat}
\end{equation}
where $a=\sqrt{1-\alpha^2-\beta^2}$. The proton $p$ state should
have isospin $I=1/2$ with $I_z=1/2$, and the neutron $n$ state
should have isospin  $I=1/2$ with $I_z=-1/2$. If the nucleon state
contains a small isospin impurities, we have
\begin{eqnarray*}
&|p\rangle=b\left|\frac{1}{2},\frac{1}{2}\right\rangle+\gamma\left|\frac{3}{2},\frac{1}{2}\right\rangle,
\\
&|n\rangle=b\left|\frac{1}{2},-\frac{1}{2}\right\rangle+\gamma\left|\frac{3}{2},-\frac{1}{2}\right\rangle,
\end{eqnarray*}
where $b=\sqrt{1-\gamma^2}$, and $\gamma$ is a very small quantity
refelcting isospin violating.\cite{isospinv1,isospinv2} The kaon
$K^+$ state should have isospin $I=1/2$ with $I_z=1/2$, and the
kaon $K^0$ state should have isospin $I=1/2$ with $I_z=-1/2$. Thus
we have
\begin{eqnarray*}
&|K^+\rangle=\left|\frac{1}{2},\frac{1}{2}\right\rangle,\\
&|K^0\rangle=\left|\frac{1}{2},-\frac{1}{2}\right\rangle.
\end{eqnarray*}

The $n K^+$ final state can be written as
\begin{equation}
|n,K^+\rangle=b\left|\frac{1}{2}-\!\frac{1}{2}\;,\;\frac{1}{2}\frac{1}{2}\right\rangle+\gamma\left|\frac{3}{2}-\!\frac{1}{2}\;,\;\frac{1}{2}\frac{1}{2}\right\rangle,
\label{fs1}
\end{equation}
and the $p K^0$ state can be written as
\begin{equation}
|p,K^0\rangle=b\left|\frac{1}{2}\frac{1}{2}\;,\;\frac{1}{2}-\;\frac{1}{2}\right\rangle+\gamma\left|\frac{3}{2}\frac{1}{2}\;,\;\frac{1}{2}-\;\frac{1}{2}\right\rangle.
\label{fs2}
\end{equation}
Isospin is conserved in hadronic decays via strong interaction,
thus the $\Theta^+$ state can not decay into $n K^+$ and $p K^0$
final states if $\Theta^+$ is a pure $I=2$ isotensor state.
Assuming that $\alpha$ and $\beta$ are of the same size as the
mixing coefficient of 0.015 estimated from the physical $\Lambda$
and $\Sigma^0$ mixing,\cite{lambdamix} a narrow width of the
$\Theta^+$ state can be understood.\cite{isotensor} We will show
in the following that the difference between $\alpha$ and $\beta$
can introduce difference between the decay probabilities of the
two decay modes for the $\Theta^+$ state.

We may expand the $\Theta^+$ state Eq.~(\ref{sstat}) by using
Eqs.~(\ref{fs1}) and (\ref{fs2}),
\begin{equation}
\begin{array}{cl}
& a|2,0\rangle + \alpha|1,0\rangle+\beta|0,0\rangle \\
\simeq & a\gamma\left|\frac{3}{2}-\!\frac{1}{2}\;,\;\frac{1}{2}\frac{1}{2}\right\rangle\left\langle\frac{3}{2}-\!\frac{1}{2}\;,\;\frac{1}{2}\frac{1}{2}\right|2,0\left\rangle\right.\\
    & {}+a\gamma\left|\frac{3}{2}\frac{1}{2}\;,\;\frac{1}{2}-\!\frac{1}{2}\right\rangle\left\langle\frac{3}{2}\frac{1}{2}\;,\;\frac{1}{2}-\!\frac{1}{2}\right|2,0\left\rangle\right. \\
    &{}+b\alpha\left|\frac{1}{2}-\!\frac{1}{2}\;,\;\frac{1}{2}\frac{1}{2}\right\rangle\left\langle\frac{1}{2}-\!\frac{1}{2}\;,\;\frac{1}{2}\frac{1}{2}\right|1,0\left\rangle\right.\\
    &{}+b\alpha\left|\frac{1}{2}\frac{1}{2}\;,\;\frac{1}{2}-\!\frac{1}{2}\right\rangle\left\langle\frac{1}{2}\frac{1}{2}\;,\;\frac{1}{2}-\!\frac{1}{2}\right|1,0\left\rangle\right.\\
    &{}+b\beta\left|\frac{1}{2}-\!\frac{1}{2}\;,\;\frac{1}{2}\frac{1}{2}\right\rangle\left\langle\frac{1}{2}-\!\frac{1}{2}\;,\;\frac{1}{2}\frac{1}{2}\right|0,0\left\rangle\right.\\
    &{}+b\beta\left|\frac{1}{2}\frac{1}{2}\;,\;\frac{1}{2}-\!\frac{1}{2}\right\rangle\left\langle\frac{1}{2}\frac{1}{2}\;,\;\frac{1}{2}-\!\frac{1}{2}\right|0,0\left\rangle\right.,
\end{array}
\end{equation}
in which the $\alpha\gamma$ term is ignored. On the other hand, we
have the Clebsch-Gorden coefficients for the isospin summation
\begin{displaymath}
\begin{array}{l@{~~}r}
\left\langle\frac{3}{2}-\!\frac{1}{2}\;,\;\frac{1}{2}\frac{1}{2}\right|2,0\left\rangle\right.=1/\sqrt{2},
&
\left\langle\frac{3}{2}\frac{1}{2}\;,\;\frac{1}{2}-\!\frac{1}{2}\right|2,0\left\rangle\right.=1/\sqrt{2}, \\
\left\langle\frac{1}{2}-\!\frac{1}{2}\;,\;\frac{1}{2}\frac{1}{2}\right|1,0\left\rangle\right.=1/\sqrt{2},
&
\left\langle\frac{1}{2}\frac{1}{2}\;,\;\frac{1}{2}-\!\frac{1}{2}\right|1,0\left\rangle\right.=1/\sqrt{2}, \\
\left\langle\frac{1}{2}-\!\frac{1}{2}\;,\;\frac{1}{2}\frac{1}{2}\right|0,0\left\rangle\right.=-1/\sqrt{2},
&
\left\langle\frac{1}{2}\frac{1}{2}\;,\;\frac{1}{2}-\!\frac{1}{2}\right|0,0\left\rangle\right.=1/\sqrt{2}.
\end{array}
\end{displaymath}
This means that
\begin{equation}
\begin{array}{cc}
a|2,0\rangle + \alpha|1,0\rangle+\beta|0,0\rangle&\simeq\frac{1}{\sqrt{2}}a\gamma\left|\frac{3}{2}-\!\frac{1}{2}\;,\;\frac{1}{2}\frac{1}{2}\right\rangle \\
&{}+\frac{1}{\sqrt{2}}b(\alpha-\beta)\left|\frac{1}{2}-\!\frac{1}{2}\;,\;\frac{1}{2}\frac{1}{2}\right\rangle \\
& {}+ \frac{1}{\sqrt{2}}a\gamma\left|\frac{3}{2}\frac{1}{2}\;,\;\frac{1}{2}-\!\frac{1}{2}\right\rangle \\
&{}+\frac{1}{\sqrt{2}}b(\alpha+\beta)\left|\frac{1}{2}\frac{1}{2}\;,\;\frac{1}{2}-\!\frac{1}{2}\right\rangle.
\end{array}
\end{equation}
The first two terms correspond to the $nK^+$ state, and the last
two terms correspond to the $pK^0$ state. The probability ratio of
the two decay modes can be expressed as
\begin{equation}
\frac{\Gamma(\Theta^+\to nK^+)}{\Gamma(\Theta^+\to
pK^0)}=\frac{\langle \Theta^+| nK^+  \rangle}{\langle \Theta^+|
pK^0  \rangle}=
\frac{(a\gamma)^2+b^2(\alpha-\beta)^2}{(a\gamma)^2+b^2(\alpha+\beta)^2}.
\end{equation}
As we mentioned before, $\gamma$ is very small and can be ignored,
we thus get
\begin{equation}
\frac{\Gamma(\Theta^+\to nK^+)}{\Gamma(\Theta^+\to
pK^0)}=\frac{(\alpha-\beta)^2}{(\alpha+\beta)^2}.
\end{equation}

We still have not taken into account of the spin and parity of
$\Theta^+$. As one of the properties of centrifugal barrier, the decay probability is directly
proportional to a factor of $k^{2L+1}$, where $k$ is the center of
mass decay momentum, and $L$ is the orbital angular momentum.\cite{gaocs}
Therefore we get the probability ratio of the two decay modes
\begin{equation}
\frac{\Gamma(\Theta^+\to nK^+)}{\Gamma(\Theta^+\to pK^0)}
=\frac{(\alpha-\beta)^2}{(\alpha+\beta)^2}(\frac{k_1}{k_2})^{2L+1},
\end{equation}
with general symmetry consideration of isospin, spin, and parity
taken into account.

In this paper we adopt the $\Theta^+$ mass $m_{\Theta^+}=
1541~\mbox{MeV}$, so we get $k_1=283.1~\mbox{MeV}$ for the
$\Theta^+ \to nK^+$ decay mode and $k_2=279.6~\mbox{MeV}$ for the
$\Theta^+ \to pK^0$ decay mode. The spin and parity of $p$ and $n$
are $J^P=\frac{1}{2}^+$, of $K^0$ and $K^+$ are $J^P=0^-$. Then
the total angular momentum of final state is $\frac{1}{2}+L$, and
parity is $(-1)^{L+1}$, where $L$ is the orbital angular momentum.
The possible spin and parity of $\Theta^+(uudd\bar{s})$ may be
$\frac{1}{2}^-, \frac{1}{2}^+, \frac{3}{2}^+, \frac{3}{2}^-,
\frac{5}{2}^-, \frac{5}{2}^+, \cdots$ . Because of parity
conservation, the corresponding orbital angular momentum of decay
modes are $0,1,1,2,2,3,\cdots$ . Higher spin states seem unlikely
for the ground state pentaquark, so we do not calculate them
further. We thus obtain  Table~1 for the probability ratio of the
two decay modes for the $\Theta^+$ state with different spin and
parity.

\begin{table}[h]
\tbl{The decay probability ratio of the two decay modes for the
$\Theta^+$ state with different spin and parity.}
{\begin{tabular}{ccc}
\hline
$ ~~J^P$ ~~& ~~L ~~& ~~$\frac{\Gamma(\Theta^+\to nK^+)}{\Gamma(\Theta^+\to pK^0)}~~$ \\
\hline
$\frac{1}{2}^-$ & 0 & $\frac{(\alpha-\beta)^2}{(\alpha+\beta)^2}\cdot1.013$ \\
\hline
$\frac{1}{2}^+$ & 1 & $\frac{(\alpha-\beta)^2}{(\alpha+\beta)^2}\cdot1.038$ \\
\hline
$\frac{3}{2}^+$ & 1 & $\frac{(\alpha-\beta)^2}{(\alpha+\beta)^2}\cdot1.038$ \\
\hline
$\frac{3}{2}^-$ & 2 & $\frac{(\alpha-\beta)^2}{(\alpha+\beta)^2}\cdot1.064$ \\
\hline
$\frac{5}{2}^-$ & 2 & $\frac{(\alpha-\beta)^2}{(\alpha+\beta)^2}\cdot1.064$ \\
\hline
$\frac{5}{2}^+$ & 3 & $\frac{(\alpha-\beta)^2}{(\alpha+\beta)^2}\cdot1.091$ \\
\hline
\end{tabular}}
\end{table}

 From Table~1, we can find that different spin and parity give
little difference to the probability ratio of the two decay modes,
as the difference between the center of mass decay momenta $k_1$
and $k_2$  is very small. The results in Table~1 are valid for the
$\Theta^+$ pentaquark to be a pure isoscalar ($I=0$) and isovector
($I=1$) state, or a mixing isoscalar and isovector state, or an
isotensor state with mixture of isoscalar and isovector
components, as they are corresponding to the specific situation of
$a=0, \alpha=0, \beta=1$ for isoscalar, or $a=0, \alpha=1,
\beta=0$ for isovector, or $a= 0$ with nonzero $\alpha$ and
$\beta$ for isoscalar and isovector mixture, or $a < 1$ with small
$\alpha$ and $\beta$ for isotensor with mixing of isoscalar and
isovector. Thus our result of the probability ratio of the two
decay modes is applicable to general situations.

 It is claimed\cite{SAPHIR} that the SAPHIR experiment suggests an isospin
of $I=0$ for the observed $\Theta^+$ state from the absence of a
signal in the $pK^+$ invariant mass distribution in the $\gamma p
\to pK^+K^-$ process. We need to examine the situation carefully,
provided that the production mechanism of $\Theta^+$ is not known
yet. In photoproduction of pentaquark $\Theta$ states on the
nucleon, we assume that the production is through the
$K^0\bar{K^0}$ and $K^+K^-$ components of the photon. The spin,
parity, and C-parity for this component of the photon should be
$J^{PC}=1^{--}$. From vector dominance model\cite{VDM} we know
that the interaction of the photon (virtual photon) can be
effectively described by the vector mesons ($V=\rho$, $\omega$,
$\phi$ with quantum numbers $J^{PC}=1^{--}$) which have both $I=0$
and $1$ states. From the available particle listing\cite{PDG} we
notice that the $I=0$ mesons ($\phi$'s) can decay into $K
\bar{K}$, whereas the $K \bar{K}$ decay ratio of $I=1$ mesons
($\rho$'s) is very small. This implies that the $I=0$ component of
the photon contains the $s\bar{s}$ content that may contribute to
$\Theta$ production via strong interaction. From this picture, the
production of $\Theta$'s in photoproduction process is through
$\gamma p \to K^0 \bar{K^0} p \to \Theta^+ \bar{K^0}$ and $\gamma
p \to K^+ K^- p \to \Theta^{++} K^{-}$ on the proton, and $\gamma
n \to K^+ K^- n \to \Theta^+ K^-$ and $\gamma n \to K^+ K^- n$
without $\Theta^{++}$ on the neutron. Thus the only process that
can contributes to $\Theta^{++}$ production is $\gamma p \to
\Theta^{++} K^-$. We assume that the $\Theta^{++}$ state, which
belongs to the isospin pentaplet
$\Theta=(\Theta^-,\Theta^0,\Theta^+,\Theta^{++},\Theta^{+++})$, is
of isospin $(I,I_z)=(2,1)$. The isospin of initial state
$\left|i\right>=\left|\gamma p\right>$ is
$\left|0,0\right>\left|1/2,1/2\right>$, whereas the isospin of
final state $\left|f\right>=\left |\Theta^{++} K^-\right>$ is
$\left|2,1\right>\left|1/2,-1/2\right>$,
 from which we find that the amplitude
$\left<f\right|\left.i\right>$ cancels exactly. However, the
amplitude of $\Theta^+$ production is also zero if $\Theta^+$ is
an isotensor;
which is in contradiction with the experimental observation. If
$\Theta^+$ and $\Theta^{++}$  are isovector partners, both the
amplitudes of $\Theta^+$ and $\Theta^{++}$ productions are
non-zero and of the same order. Thus the observation of no
$\Theta^{++}$ peak in the photoproduction on the nucleon can
exclude both the $I=1$ and $I=2$ cases, if the mechanism of
$\Theta$ production is through the isoscalar component of the
photon as we suggested.

If the $I=1$ component of electromagnetic current (i.e., the
photon) contributes dominantly to the production of pentaquark
$\Theta$'s, as some literature assumed,\cite{isotensor,SAPHIR} we
now show that the absence of no $\Theta^{++}$ peak cannot rule out
an isovector $\Theta^+$. We assume that the $\Theta^{++}$ state,
which belongs to the isospin triplet
$\Theta=(\Theta^0,\Theta^+,\Theta^{++})$, is of isospin
$(I,I_z)=(1,1)$. The isospin of initial state
$\left|i\right>=\left|\gamma p\right>$ is
$\left|1,0\right>\left|1/2,1/2\right>$, whereas the isospin of
final state $\left|f\right>=\left |\Theta^{++} K^-\right>$ is
$\left|1,1\right>\left|1/2,-1/2\right>$. From Clebsch-Gordan
coefficients we have
\begin{equation}
\begin{array}{l}
\left|i\right>=\left|1,0\right>\left|\frac{1}{2},\frac{1}{2}\right>=\sqrt{\frac{2}{3}}\left|\frac{3}{2},\frac{1}{2}\right>
-\sqrt{\frac{1}{3}}\left|\frac{1}{2},\frac{1}{2}\right>; \\
\left|f\right>=\left|1,1\right>\left|\frac{1}{2},-\frac{1}{2}\right>
=\sqrt{\frac{1}{3}}\left|\frac{3}{2},\frac{1}{2}\right>+\sqrt{\frac{2}{3}}\left|\frac{1}{2},\frac{1}{2}\right>,
\end{array}
\end{equation}
 from which we find that the amplitude
$\left<f\right|\left.i\right>$ cancels exactly. Thus the
observation of no $\Theta^{++}$ peak in the photoproduction on the
nucleon can not rule out the $I=1$ case for $\Theta^+$, but the
$I=2$ case in this situation. The pentaquark $\Theta^+$ state
could be thus pure $I=0$ state, or $I=0$ state with small mixture
of $I=1$ component, or $I=1$ state with small mixture of $I=0$
component if the isovector part of the photon contributes
dominantly to the $\Theta^+$ production. Therefore our results of
decay probability ratio may provide useful information to
constrain the isospin of the $\Theta^+$ state.

Though Nature permits the existence of isospin mixing state (for
example, the photon is an $I=0,1$ mixing state), approximate
isospin conservation in the strong interactions implies that the
mass matrix of the strongly interacting particles will
approximately commute with the isospin generators, and thus that
the mass eigenstates will be approximate isospin eigenstates. This
means that the most likely possibilities should be $\alpha \approx
1$ with small $\beta$, or $\beta \approx 1$ with small $\alpha$ if
the production mechanism of $\Theta^+$ is through the $I=1$
component of photon as most literature suggested. From our above
argument for an important isoscalar contribution of the photon to
$\Theta$ production, we suggest that the observed $\Theta^+$ is
likely an isoscalar.
It is thus necessary to check whether our results
are still experimentally measurable for the case of a large
$\alpha \approx 1$ with small $\beta$, or vise verse. Assuming
that $\alpha=0.98$ and $\beta=0.02$, which is of the right order
of isospin breaking for baryons, we find that
$\frac{(\alpha-\beta)^2}{(\alpha+\beta)^2}\approx 0.92$, which is
measurable within experimental precision provided with enough
number of events in the near future. In case of a larger isospin
breaking, say that $\beta \approx 0.05$, we find that
$\frac{(\alpha-\beta)^2}{(\alpha+\beta)^2}\approx 0.81$. This
implies that our results are also effective to reveal isospin
breaking effect in $\Theta^+$ decays, if $\Theta^+$ is an
isoscalar as most authors predicted.

 It is necessary to point out that the decay ratio we discussed is
based on an assumption of isospin conservation in the decay
processes. There is also a suggestion\cite{BM03} that the observed
$\Theta^+$ state might be a heptaquark with the overlap of a pion,
a kaon, and a nucleon. In this case, the decay ratio of two decay
modes may differ from the prediction in our work. It is also
possible that the decays are not through strong interaction, or
other isospin violating mechanism may involve. The decay ratio
predicted in this work is thus helpful to reveal detailed
properties of the observed $\Theta^+$ state, and to confirm the
observed $\Theta^+$ state as being of pentaquark $uudd\bar{s}$
configuration. From another point of view, there may be higher
$\Theta^{*+}(uudd\bar{s})$ state beyond the anti-decuplet
$\{\bar{10}\}$ multiplet.\cite{WK03,Wu} It has been speculated
from QCD sum rules that the isoscalar, isovector, and isotensor
states of $\Theta^+$ and $\Theta^{*+}$ pentaquarks may lie close
to each other.\cite{Zhu} The results in this work are applicable
to these $\Theta^{*+}(uudd\bar{s})$ states if the mass difference
is considered.

 In summary, the decay probability ratio
of the two decay modes $\Theta^+ \to nK^+$ and $\Theta^+\to pK^0$
for the pentaquark $\Theta^+$ state is examined with general
symmetry consideration of isospin, spin, and parity. This ratio
depends on two coefficients $\alpha$ and $\beta$ which can
represent a pure isoscalar or isovector state, or an isoscalar and
isovector mixing state, or an isotensor state with mixture of
isoscalar and isovector components. The dependence on spin and
parity of the pentaquark $\Theta^+$ state is also considered, and
the effect is found to be small due to small difference between
the center of mass decay momenta of the two decay modes. We also
provided an analysis on the constraint of the isospin of
$\Theta^+$ from the absence a peak in the $pK^+$ invariant mass
distribution in the $\gamma p \to pK^+K^-$ process. Future
experimental results about the decay probability ratio may confirm
and provide information about the properties of the pentaquark
$\Theta^+$ state.

\section*{Acknowledgements}
We acknowledge the helpful discussions with Kuang-Ta Chao, Yanlin
Ye, Chuan Liu, Xiaorui Lu, Weilin Yu, Shi-Lin Zhu, K.~Maltman, and
H.E. Jackson. This work is partially supported by National Natural
Science Foundation of China under Grant Numbers 10025523, 10145008
and 90103007.


\begin{thebibliography}{99}
\bibitem{footnote}
In CLAS experiment, an additional $p$ is also measured in the
exclusive $\gamma D\to K^- p (\Theta^+ \to K^+ n)$ process, just
as the $\gamma^* D \to K^- p \Theta^+$ process suggested in
Ref.~\refcite{GM99}. 
\end{thebibliography}
\end{document}